# Spin polarized diamond neutral vacancy

S. Sheikholeslami and A. Shafiekhani

*Department of Physics, Alzahra University, Tehran, 19938-91167, Iran*

**Abstract**

The atomic orbital basis and atomic pseudopotential for electron-ion interaction in frame work of density function theory used to calculate $C_{70}H_{84}$ spin polarized diamond cluster with neutral vacancy. The band gap has been found to be 1.70 eV. The chosen cluster comprises electronic properties depending on the type of its spin. It could be used as a data transporter. The calculated band gap, density of state and spin properties illustrate how beneficial it could be in spintronic devices.

## Introduction

Natural diamonds are classified as types Ia, Ib, IIa and IIb by the impurities that are found within them [1]. The absorptions in natural diamond are due to vacancies throughout their bulk, which is old but still an interesting subject.

The vacancies in diamonds could be experimentally produced by sufficient energy irradiation with, for example; electrons, neutrons, alpha-particles and gamma-rays, to displace some of the carbon atoms [2-7]. The vacancies in diamond have the same tetrahedral ($T_d$) space symmetry as diamond.

The defect-free materials generally are nonmagnetic while local magnetic moment is induced around the vacancy sites and interacts with each other to cause the material to become magnetic. Such magnetism called intrinsic because no foreign species are employed, thus this type of magnetism has attracted attention as a new way to synthesize dilute magnetic semiconductors [8].

The simplest and the most fundamental defects in semiconductors are vacancies. These defects and impurities have related impacts on electrical and optical properties of semiconductor materials [9,10]. Study on the lattice vacancy is first reasonable step toward understanding of added impurities in general. Recently, more attention on color centers in diamond, i.e. carbon vacancy and nitrogen-vacancy centers have been nominated as a practical candidate for application in quantum information processing and quantum cryptography [11,12].

There exist two main theoretical approaches in this subject; molecular and large cluster models [13]. Both approaches have their own advantages and in some cases, are accomplishing each other. The configuration interaction (CI) method is employed in molecular models to calculate electronic structure of the vacancies. In addition, the role of e–e interaction is in more consideration [14]. These models appear from the old computational design of the CI, which has low computational efficiency.

Deficiency of the CI formalism have imposed the molecular models towards using a more approximated form of the Hamiltonian, which neglect e–e exchange terms and employed proper excitation energies, empirically [15-17]. The empirical models explain new experimental results, dipole transition intensities and electron paramagnetic resonance (EPR) measurement, on the neutral vacancy (V°) in diamond and negatively charged (V ) vacancies [15,17]. With all attempts to improve the CI computational scheme, it has little success [15].

The well-developed computational scheme of density functional theory (DFT) [18] and diffusion Monte-Carlo (DMC) theory [19] employed to explain the experimental data.

The vacancy in diamond has many challenged, and many approaches were investigated in which emphasize was placed on the benefit of using a large cluster [13]. The cluster model has more consideration to the ground-state properties and coupling of the defect to the surrounding bulk [20]. In addition, first principle DMC calculation was reported for the similar excitation (GR1) in the neutral vacancy in diamond [19]. Although significant progress in calculations of optical excitations by *ab initio* theories, to explain EPR results [18] has occurred, using molecular approach have recommended [15,17]. Introducing a generalized Hubbard model with one semi empirical parameter, simultaneously GR1 and ND1 transition energies obtained more accurately [20].

In this paper, we use the density functional theory (DFT) based on the Generalized-Gradient-Approximation (GGA) method to calculate the density of states (DOS) of neutral (V°) in diamond by means of investigating their spin properties, band gap and electronic properties.

**Computational method and modeling**

We used SIESTA-2.0.2 package [21] on the base of DFT method to calculate V° in diamond. The atomic pseudopotentials are adopted for electron-ion interaction. We also, used DZ polarization (DZP) orbitals. The GGA by Perdew-Burke-Ernzerhof (PBE) [22] for the exchange-correlation potentials is employed. The coordinate optimization by conjugate gradients (CG) algorithms was chosen for the geometry optimization. In all the calculations, the wave functions are expanded into plane waves up to a cutoff energy of 250 Ry. We chose Hamiltonian diagonalization method to solve the cluster model [21]. The 4×4×4 K grid was created for the Monkhorst-Pack scheme. The optimization of the atomic coordinates continued until the maximum total force become to 0.04 eV/Å.

The diamond cluster $C_{70}H_{84}$ with a neutral vacancy has 70 carbons which are saturated with hydrogen [23,20,24]. The cluster is spin polarized and about 8 nm in diameter. We allowed all atoms of the cluster to relax freely during the calculations. To calculate the cluster, a 50 Å simple cubic box was employed to isolate the cluster in vacuum.

**Results and discussion**

The result of the calculation for the electronic state of the V° is summarized in Fig. 1. For the V°, as is shown in Fig. 1, a dipole-allowed transition from a doubly degenerate ground state to a triply degenerate excited state which is spin singlet states, $^1E \to {^1T}$ has calculated.

We found that the density of states at the Fermi level is 0.23 (eV/atom). The HOMO is separated from the LUMO by a gap of 1.70 eV, which is in agreement with the experimental value of 1.67 eV [25] for the GR1 transition energy.

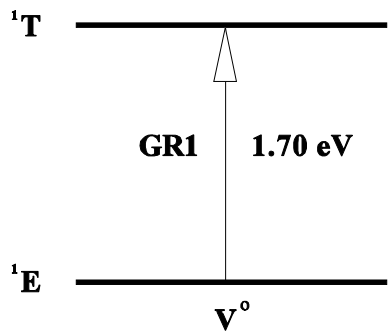

Figure 1: Transition of $^1E \rightarrow {}^1T$ state for $V°$.

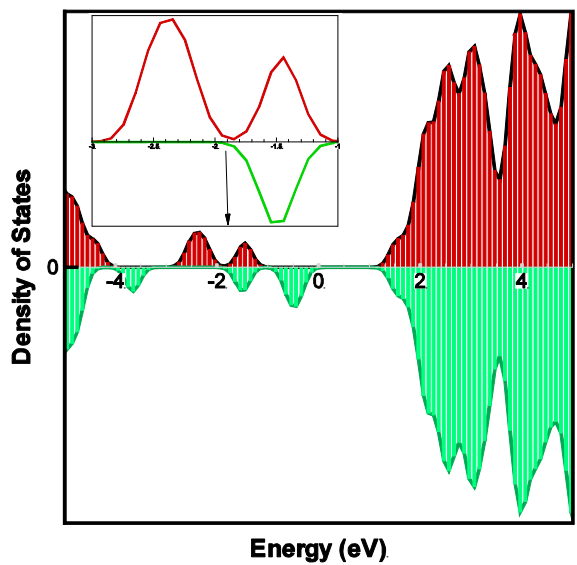

Figure 2: Total density of majority (↑) and minority (↓) spin states.

The total density of majority (↑) and minority (↓) spin states are shown in Fig. 2. The DOS of the minority spin states has a band gap and the Fermi level is close to the conduction band edge. The density of majority spin states is almost 20% at the Fermi level. Accordingly, the spin states at $E_F$ are partially polarized. In other word, majority spin band plays roll as semi-metal but minority spin bands as semiconductor.

## Conclusion

Using *ab initio* density function theory for total energy calculations, we evaluated the density of states at the Fermi level. Also, the band gap of spin polarized in diamond cluster, $C_{70}H_{84}$, with neutral vacancy has been found to be 1.70 eV, which in compare with others, is in good agreement with the experimental value of 1.67 eV.
Our results suggest that using an atomic orbital basis instead of molecular orbitals is more convenient to explain the physical properties of the system.
We further find that the majority spin band play roll as semi-metal but minority spin bands as semiconductor in spin polarized diamond cluster $C_{70}H_{84}$ with neutral vacancy.

## Acknowledgement
The authors would like to thank school of Nanoscience, Institute for Research in Fundamental Sciences (IPM) for providing computing server and VC for Research of Alzahra University for financial support.

**References**
[1] H. J. Lim, S. Y. Park and H. S. Cheong, J. Korean Phys. Soc. **48**, 1556 (2006).
[2] B. Campbell and A. Mainwood, Phy. Stat. Sol. (a) 181, 99 (2000).
[3] A. Mainwood, J. Cunningham and D. Usher, Mater. Sci. Forum **787**, 258 (1997).
[4] K. H. Han, J. Korean Phys. Soc. **48**, 1427 (2006).
[5] Y. K. Lim, B. S. Park, S. K. Lee and K. R. Kim, J. Korean Phys. Soc. **48**, 777 (2006).
[6] Y. S. Kim, E. S. Choi, W. S. Kwak and Y. J. Shin, J. Korean Phys. Soc. **51**, 503 (2007).
[7] J. Park, S. J. Shin and M. J. Seong, J. Korean Phys. Soc.**53**, 2312 (2008).
[8] Y. Yang, O. Sugino, and T. Ohno, Phys. Rev. **B 85**, 035204 (2012).
[9] J. P. Goss, B. J. Coomer, R. Jones, C. J. Fall, P. R. Briddon, et al, Phys. Rev. **B 67**, 165208 (2003).
[10] C. Glover, M. E. Newton, P. Martineau, D. J. Twitchen, and J. M. Baker, Phys. Rev. Lett. **90**, 185507 (2003).
[11] A. Beveratos, R. Brouri, T. Gacoin, A. Villing, J. P. Poizat, et al, Phys. Rev. Lett. **89**, 187901(2002).
[12] P. R. Hemmer, A. V. Turukhin, M. S. Shahriar, and J. A. Musser, Opt. Lett. **26**, 361 (2001).
[13] G. Davies and N. B. Manson, in: Properties and Growth of Diamond, edited by G. Davies (IEE, London, 1994), p. 159.
[14] C. A. Coulson and M. J. Kearsley, Proc. R. Soc. **241**, 433 (1957).
[15] A. Mainwood and A. M. Stoneham, J. Phys.: Condens. Matter **9**, 2453 (1997).
[16] M. Lannoo and J. Bourgoin, in: Point Defects in Semiconductors, I: Theoretical Aspects (Springer, Berlin, 1981), p. 141.
[17] J. E. Lowther, Phys. Rev. **B 48**, 11592 (1993).
[18] J. A. van Wyk, O. D. Tucker, M. E. Newton, J. M. Baker, G. S. Wood, et al, Phys. Rev. B 52, 12657(1995).
[19] R. Q. Hood, P. R. C. Kent, R. J. Needs, and P. R. Briddon, Phys. Rev. Lett. **91**, 076403-1 (2003).
[20] M. Heidari Saani , M. A. Vesaghi, K. Esfarjani, and A. Shafiekhani, Phys. Stat. Sol. (b) **243**, 1269–1275 (2006) / DOI 10.1002/pssb.200541032**.**
[21] E. Artacho, J. D. Gale, A. Garcia, J. Junquera, P. Ordejon, et al, J. M. Soler (1996-2008).
[22]J. P. Perdew, K. Burke, M. Ernzerhof, Phys. Rev. Lett. 77, 3865 (1996).
[23] S. J. Breuer, P. R. Briddon, Phys. Rev. B **51**, 6984 (1995).
[24] A. Zywietz, J. Furthmuller, F. Bechstedt, Phys. Stat. Sol. (b) **210**, 13 (1999).
[25] J. Walker, Rep. Prog. Phys. **42**, 1605 (1978).